\documentclass{jaa}
\usepackage{natbib}
\usepackage{graphicx}
\usepackage{xcolor}
\usepackage{hyperref}
\usepackage{subfigure}
\hypersetup{colorlinks=true,linkcolor=blue,citecolor=blue,filecolor=blue,urlcolor=blue,}
\newcommand{\angstrom}{\textup{\AA}}

\begin{document}\sloppy

\title{Abundance Stratification in Type Iax SN~2020rea with TARDIS}

\author{Sohini Kayal\textsuperscript{1,*}, P. Gayatri\textsuperscript{2,*}, Mridweeka Singh\textsuperscript{3}, and Kuntal Misra\textsuperscript{4}}
\affilOne{\textsuperscript{1}Department of Physics and Department of Electrical \& Electronics Engineering, Birla Institute of Technology \& Science - Pilani, Hyderabad, 500078, India.\\}
\affilTwo{\textsuperscript{2}School of Physical Sciences, National Institute of Science Education and Research, An OCC of Homi Bhabha National Institute, Bhubaneswar, 752050, India.\\}
\affilThree{\textsuperscript{3}Indian Institute of Astrophysics, II Block, Koramangala, Bengaluru, 560034, India.\\}
\affilFour{\textsuperscript{4}Aryabhatta Research Institute of Observational Sciences, Manora Peak, Nainital, 263001, India.}

\twocolumn[{

\maketitle

\coauthors{}
\corres{\textsuperscript{1}kayal.sohini@gmail.com, \textsuperscript{2}gaya3p13@gmail.com}

\msinfo{19 May 2025}{30 December 2025}

\begin{abstract}

Using the 1D Monte Carlo–based radiative transfer code \textsc{TARDIS}, we investigate the spectral evolution of the Type Iax supernova (SN) 2020rea from  $-$7 days before to +21 days after maximum light. Our best-fit models indicate stratified, velocity-dependent abundances at early times, successfully reproducing most observed spectral features. As the SN evolves, the ejecta transition from a layered to a more homogeneous composition, posing an alternative to pure deflagration models that predict fully mixed ejecta. These results highlight the need for further investigation, as current pure deflagration models cannot fully explain the origin or spectral properties of Type Iax SNe like SN 2020rea.

\end{abstract}

\keywords{techniques: spectroscopic -- supernova modelling -- general -– supernovae: individual: SN~2020rea -– galaxies: individual: UGC~10655.}
}]

\doinum{https://doi.org/10.1007/s12036-026-10136-5}
\artcitid{19}
\volnum{47}
\year{2026}
\pgrange{1--10}
\setcounter{page}{1}
\lp{10}

\section{Introduction}

Type Iax supernovae (SNe Iax) constitute a distinct subclass of Type Ia supernovae (SNe Ia), exhibiting systematically lower luminosities and reduced explosion energies \citep{li_sn_2003, foley_type_2013}. They possess lower velocities (${\sim}$ 2000 - 7000 km/s, \citealt{alsabti_type_2017}) as compared to the typical expansion velocities of SNe Ia ($\sim $10,000 km/s, \citealt{folatelli_spectroscopy_2013}) at maximum light. They exhibit a diverse range of peak magnitudes, from $M_r = -12.7$ mag (SN 2021fcg, \citealt{karambelkar_faintest_2021}) to $M_r = -18.3$ mag (SN 2012Z, \citealt{mccully_luminous_2014}). Due to their observational similarities, these SNe are grouped under the 2002cx-like objects, whose first identified member was the prototype SN~2002cx \citep{li_sn_2003}. 

Brighter SNe Iax display a homogeneous spectral evolution, similar to SNe Ia post maximum brightness (except with lower line velocities, \citealt{jha_late-time_2006}) while fainter type Iax SNe evolve more rapidly in velocity.
The pre-maximum spectra show intermediate-mass (Si, S, and Ca) and iron-group elements (IMEs and IGEs), similar to SNe Ia. This similarity also extends to the near-UV region. Likewise, in their near-infrared spectra at maximum light, SNe Iax show Fe II and Si III lines similar to SNe Ia, except with lower line velocities \citep{alsabti_type_2017}. 

However, the late-time spectra of SNe Iax are very different from SNe Ia. Unlike SNe Ia, SNe Iax never truly enter a fully \textit{nebular} phase in which broad forbidden lines dominate the optical spectrum \citep{mccully_luminous_2014,stritzinger_comprehensive_2015}. SNe Iax show permitted lines of Fe II (with velocities $<$ 2000 km s$^{-1}$) along with forbidden lines of Ca II, Fe II and Ni II (often with lower line widths $<$ 500 km s$^{-1}$; \citealt{foley_late-time_2016}, \citealt{alsabti_type_2017}). The relative strengths of these lines vary significantly among different members of SNe Iax \citep{foley_type_2013}. The very late time spectra of SNe Iax show very slow evolution, from around +200 to +400 days after maximum light. However, it has been observed that the line velocities unusually dip after +450 days, suggested to be caused by a weakening of the super-Eddington wind driven by a bound remnant \citep{foley_late-time_2016, camacho-neves_over_2023}.

These similarities and differences between SNe Ia and SNe Iax have been studied in depth, but their possible progenitors and explosion models are still being investigated. Due to the low luminosity and the less energetic nature of SNe Iax, two of the leading ideas are that of a carbon-oxygen (CO, \citealt{fink_three-dimensional_2014}) or a hybrid carbon-oxygen-neon (CONe; \citealt{meng-2014}, \citealt{kromer-2015}) white dwarf (WD) undergoing incomplete deflagration. Helium has also been detected in some SNe Iax spectra, which could be due to a WD interaction with a He-burning star companion \citep{foley_type_2013}. This deflagration is predicted not to completely disrupt the star \citep{Kromer_2012, magee_analysis_2021}. However, certain questions related to this explanation require further investigation. One of these is to understand the structure of the ejecta, which has been predicted to be homogeneous in nature \citep{magee_analysis_2021}. Preliminary spectroscopic analysis suggests that a strong mixing of elements exists in the ejecta. However, this might not be the case on further inspection of the spectral models in the earlier epochs (pre-maximum), as the pure deflagration model cannot explain certain spectral deviations.

To predict the nature of the SN and its elemental abundances, we use the 1D radiative transfer code \textsc{tardis} \citep{kerzendorf_spectral_2014}. \textsc{tardis} has been used extensively in literature to investigate a diverse range of SNe Iax \citep{magee_type_2016, barna_type_2018, barna_sn_2020, camacho-neves_over_2023}. Using specific input parameters, \textsc{tardis} generates synthetic spectra at each epoch using Monte Carlo simulations, which are then fine-tuned as we compare them with the observed spectra. Thus, we can map the entire spectral evolution at different epochs and study different regions of the SN ejecta. 
In this paper, we present a detailed spectral analysis of SN~2020rea, where we compare our findings with the results of the previous spectral model \citep{singh_optical_2022}, which reasons that the pure deflagration of a Chandrasekhar-mass WD is the most promising explosion scenario for the SN.
In section~\ref{target of study}, we summarise the properties of SN~2020rea. Section~\ref{section_tardis} describes the spectral modelling radiative transfer code \textsc{tardis}, followed by a re-analysis of SN~2020rea using the uniform abundance model in section~\ref{unifrom}. The uniform abundance modelling in \cite{singh_optical_2022} and this work does not fully explain certain observational spectral features in SN~2020rea. We, therefore, use the stratified abundance model \citep{Stehle_Mazzali_Benetti_Hillebrandt_2005, barna_abundance_2017} and discuss the results in section~\ref{stratified}. Finally, section \ref{conclusion} presents a summary of our findings in this paper.

\section{Target of Study: SN~2020rea} \label{target of study}

SN~2020rea was discovered on August 11, 2020 (JD= 2459072.702), in the host galaxy UGC~10655 at a redshift of 0.02869$\pm$0.00015 \citep{falco_updated_1999}, by the Supernova and Gravitational Lenses Follow-up (SGLF) team in the Zwicky Transient Facility (ZTF) data \citep{Perez-Fournon2020}. The SN was classified as Type Ia-pec based on the spectroscopic features \cite{Poidevin_Perez2020}. SN~2020rea has been thoroughly analysed in \cite{singh_optical_2022}. They estimated the explosion epoch by fitting a radiation diffusion model to the pseudo-bolometric light curve, and it was found to be August 9, 2020 (JD =  $2459070.64^{+1.45}_{-0.76}$). The SN lies at the brighter end of the SNe Iax luminosity distribution ($M_\text{V, peak} \sim -18.30\pm 0.12$ mag), similar to SNe~2011ay \citep{szalai_early_2015} and 2012Z \citep{stritzinger_comprehensive_2015, yamanaka_oister_2015}. The peak absolute magnitude in $g$-band is $M_\text{g, peak} \sim -17.34 \pm 0.03$ mag, and occurs on JD 2459084.74. Throughout this study, we have used the g-band maximum ($g_\text{max}$) as a reference for further analysis. Table~\ref{SN_details} lists other details of the SN and its host galaxy derived in \cite{singh_optical_2022}.

\begin{table}
    \centering
    \caption{Details of SN~2020rea adapted from \cite{singh_optical_2022}.}
    \begin{tabular}{cc}
    \hline
        RA (J2000.0) & $16^h59^m37.82^s$\\
        Dec. (J2000.0) & $56^\circ04'08.48''$\\
        Galactic extinction {\it E(B-V)} & 0.02 mag\\ 
        Host extinction {\it E(B-V)} & 0.08 mag\\ 
        Host galaxy & UGC~10655\\
        Redshift & 0.02869$\pm$0.00015 \\
    \hline
    \end{tabular}
    \label{SN_details}
\end{table}

Spectroscopic observations of SN~2020rea were conducted $\sim$ 5 days after discovery and continued for $\sim$ a month, using the FLOYDS spectrograph on the 2m FTN telescopes. The spectrograph provides a wavelength range of 3300--11000 \angstrom\, with resolution ranging between 400 and 700. The spectra were subsequently reduced as described in \cite{singh_optical_2022}. 

\cite{singh_optical_2022} performed an in-depth spectral analysis and studied the spectral evolution. They compared SN~2020rea with other members of the SNe Iax class. In the pre-maximum spectra, the Fe III feature between 4000 and 5000 \angstrom\ was prominently seen. These epochs showed well-developed P-Cygni profiles with relatively broad absorption features. They also showed Ca II and Si II features in the blue region, Si III, S II, Fe III, and a relatively weak Si II feature around 6000 \angstrom. These spectroscopic features are typical of brighter SNe Iax, including SNe~2011ay and 2012Z. Si II, Fe II, and Fe III features were all prominently seen in the near-maximum and maximum spectra. The post-maximum spectra of SN~2020rea showed a relatively weak Ca II triplet feature around 8500 \angstrom. The Fe II multiples, Fe III, and Cr II lines were also prominently visible in these spectra. The +20.9 day spectrum closely resembled the +20 day spectrum of SN~2012Z.

\cite{singh_optical_2022} also performed spectroscopic modelling for three epochs: $-$4.0, 0.0 and +9.9 days using the uniform abundance model (see section~\ref{section_tardis} for details) in \textsc{tardis} \citep{kerzendorf_spectral_2014}. The photospheric velocity required for the modelling was derived from the Si II line at 6355 \angstrom\, and was varied between 6000 and 6800 km s$^{-1}$. Due to the degeneracy in the parameters used in \textsc{tardis} fit, the spectral models are not necessarily unique. The overall continuum matches well with the observed spectrum and Fe features between 4000 and 5000 \angstrom\, were well reproduced. In the modelled spectrum around maximum brightness, the region between 4000 and 5200 \angstrom\, matched the observed spectrum well. The observed spectral features of the +9.9 day spectrum were well reproduced by the model with a significant amount of IGEs.

Since the uniform abundance model provides a good fit for the +9.9 day spectrum, this indicates a well-mixed ejecta in the later phases of the SN explosion, as is expected in a deflagration scenario \citep{gamezo_thermonuclear_2003}. However, for the pre-maximum spectrum, the model did not produce a good fit, especially near 5000 \angstrom\, and the ``W'' feature at $\sim$ 6000 \angstrom\ was not reproduced. 
The origin of this line has been attributed to the Sulphur (S) during the early evolutionary stages, which is then defined by Fe as the SN transitions into a Fe-dominated phase. Although the model reproduces most of the observed spectral features, it produces an inferior fit in the 3400--4000 \angstrom\,wavelength range for the maximum and pre-maximum spectra.

\section{Spectral Modelling}

\subsection{Radiative Transfer Code - TARDIS} 
\label{section_tardis}

\textsc{tardis} (v2024.12.15) is an open-source Monte Carlo radiative-transfer spectral synthesis code for 1D modelling of SN ejecta \citep{kerzendorf_spectral_2014,kerzendorf_tardis-sntardis_2023}. \textsc{tardis} assumes a spherically symmetric and homogeneously expanding ejecta around an opaque core emitting a blackbody continuum. Photon packets are tracked as they move from the photosphere through the ejecta, which are then used to calculate the emerging spectrum. As shell velocities do not change over time, the density is written as a function of the ejecta velocity. The density is uniform throughout the shell and is determined by the velocity/radius at the center, using Eq. \ref{density_eq}.

\begin{equation} \label{density_eq}
\rho\left(v, t_{\text {exp}}\right)=\rho_0\,e^{-v / v_0}\left(\frac{t_0}{t_{\text {exp }}}\right)^3
\end{equation}

Model fittings are usually performed by changing only the time-dependent parameters – time since explosion ($t_\text{exp}$), luminosity, inner boundary of the modelling volume and mass fractions of $^{56}$Ni and $^{56}$Co. The radius of the inner boundary ($r_\text{inner}$) is calculated from the velocity of the inner boundary ($v_\text{inner}$) using the relation $r_\text{inner}=v_\text{inner}\cdot t_\text{exp}$. Other parameters include the density profile, velocity of the outer boundary, elemental abundances of C, O, Na, Mg, Si, S, Ca, Cr, Fe, Co, $^{56}$Ni and plasma properties of the ejecta, which follow a somewhat consistent pattern for all SNe Iax \citep{barna_type_2018}. 

The bolometric luminosity of SN~2020rea is calculated using Supernova Bolometric Lightcurves \citep[\textsc{SuperBol};][]{Nicholl_2018} from the observed photometric magnitudes. The template abundance profile defined in \cite{barna_type_2018} was used as a starting point and later varied following the properties of SN~2020rea. To find the velocity of the inner boundary ($v_\text{phot}$), a Markov-Chain Monte Carlo (MCMC) fit was performed using the \textsc{emcee} package \citep{foreman-mackey_emcee_2013} with the Si II absorption line (6355 \angstrom) as the baseline \citep{blondin_using_2006}. The method uses least-squares approximation to find the best-fitting model parameters of the relevant absorption line. These parameters are then used as `guess' values for the  MCMC fit to the absorption line, which determines the required parameters with full posterior distributions.

In addition to the spectral modelling at $-4$.0, 0.0, and +9.9 days in \cite{singh_optical_2022}, we have performed modelling of the entire spectral time series, presented in \cite{singh_optical_2022}, from $-$7 to +20.9 days. {Model settings are provided in Table \ref{table_settings}.

\subsection{Uniform Abundance model} \label{unifrom}

\begin{figure*}
    \includegraphics[width=\textwidth]{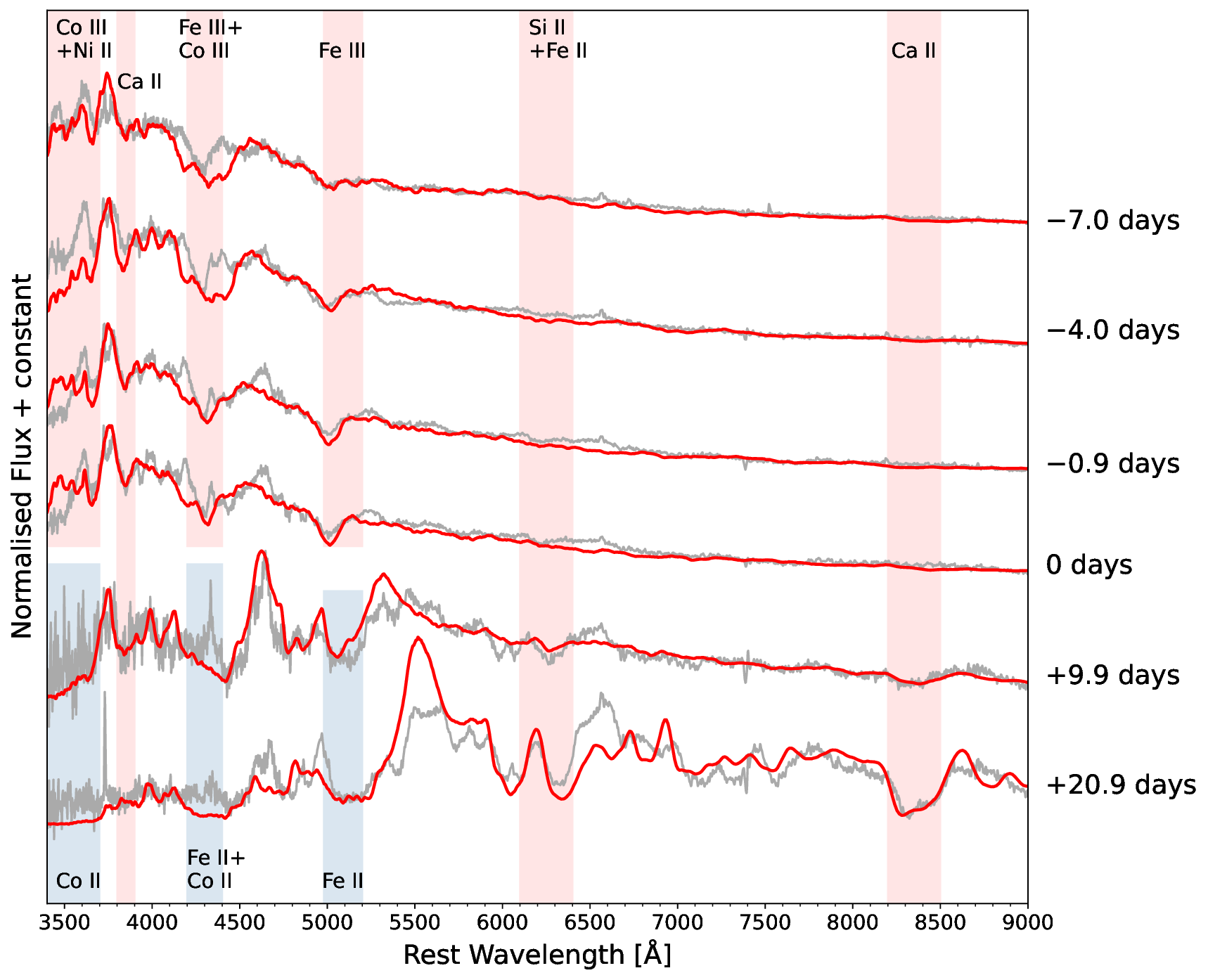}
    \caption{Observed (grey) and modelled (red) spectra for six different epochs of SN~2020rea, assuming a uniform abundance profile. The marked lines were identified by the \textsc{TARDIS} line identification tool. These were compared with \cite{li_sn_2003} and \cite{barna_type_2018}. Fe III and Co III lines in the earlier epochs transform into their corresponding lower ionisation states as the SN evolves.}
    \label{uniform_spectra}
\end{figure*}

For a more detailed re-analysis of SN~2020rea, spectral modelling was performed for all six epochs from $-$7.0 days to +20.9 days, assuming uniform elemental abundance throughout the ejecta for each epoch. The parameters used are defined in Table \ref{uniform_parameters}. The best-fit model for each epoch was defined by varying the mass fractions of elements present in the ejecta. Since accurate fitting methods for spectra are difficult, the criteria for best fit were defined mainly by matching the peaks and dips of relevant elements in the ejecta. Additional, an exponential density profile was used (Eq.~\ref{density_eq}) with $t_0=2$ days, $\rho_0 = 6 \times 10^{-11}$ g cm$^{-3}$, and $v_0=7000$ km s$^{-1}$.

\begin{table}
\centering
\caption{Model Parameters used throughout all the TARDIS models.}
\begin{tabular}{cc}
\hline
    Setting & Value \\ \hline
    Electron Scattering & Enabled \\
    Ionization & \begin{tabular}[c]{@{}c@{}}Local Thermodynamic\\Equilibrium (LTE)\end{tabular} \\
    Excitation & Dilute LTE \\
    Radiative rates type & Dilute Blackbody\\
    Line interaction type & Scatter \\
    No. of packets & $4\times 10^4$\\
    No. of virtual packets & 10\\
    Iterations & 30\\
    \hline
\end{tabular}
\label{table_settings}
\end{table}

As seen in Figure~\ref{uniform_spectra}, the observations and the generated \textsc{tardis} model spectra are in good agreement. The model also fits well with the overall continuum. This pure deflagration model fits the main features and explains most of the observational attributes of SN~2020rea. 

The abundance models described in \cite{singh_optical_2022} generated synthetic spectra that broadly match the observed spectral features after 4000 \angstrom, but fail to reproduce those from 3400--4000 \angstrom. This was resolved by assuming zero Cr in our uniform abundance profile. Flux suppression due to Cr II near 3800 \angstrom\,has previously been observed in \cite{barna_abundance_2017}. No strong features were observed due to C and O in any epoch.

Our model reproduced significant peaks around $\sim$ 4000 \angstrom\, which can be affiliated to Ca II \citep{blondin_using_2006}. The observed and synthetic pre-maximum and maximum spectra, show a significant dip at $\sim$ 4300 \angstrom\, which can be attributed to Co III.  However, for near maximum brightness, the observed peak around 4700 \angstrom\ has not been reproduced by our uniform abundance model. The Fe III feature at $\sim$ 5200 \angstrom\, which grows stronger as the SN evolves from pre-maximum to post-maximum, can also be seen in the synthetic spectra.

For the post-maximum spectra, our model could not replicate the observed peaks $\sim$ 4300 \angstrom. The model also could not accurately reproduce the ``W'' feature at $\sim$ 6000 \angstrom, which grows more prominent throughout the epochs. In the +20.9 day spectrum, spectral features around 5500 \angstrom\, and 6500 \angstrom\, generated by our model show a significant deviation from the observed spectrum.

The post-maximum spectra also show an increasingly strong Ca II line around 8350 \angstrom, which the model reproduced. These spectra were fitted with a higher mass-fraction of IGEs, which signifies the inner region of the ejecta being IGE rich. Higher amounts of O and Ne were also used as filler elements for the post-maximum spectra.

\begin{table}
    \centering
    \caption{Parameters used for spectral modelling for the respective epochs} 
    \begin{tabular}{cccc}
    \hline
        Date & Phase$^{a}$ & Luminosity & Velocity \\ 
         & (days) & ($\log L/L_\odot$) & (km s$^{-1}$)\\ 
    \hline
        16-08-2020 & $-$7.0 & 8.75 & 6900\\
        19-08-2020 & $-$4.0 & 8.90 & 6800\\
        22-08-2020 & $-$0.9 & 9.12 & 6600\\
        23-08-2020 & 0 & 9.15 & 6500\\
        02-09-2020 & +9.9 & 8.99 & 6000\\
        13-09-2020 & +20.9 & 8.80 & 5800 \\
    \hline
    \multicolumn{4}{l}{$^{a}$w.r.t. $g_\text{max}$}
    \end{tabular}
    \label{uniform_parameters}
\end{table}

This leads us to try a different model to better describe the observed spectra. As explained in section~\ref{stratified}, the stratified model varies the mass fractions of elements throughout the shells to describe a state of ejecta that is not fully mixed. Using this, we attempt to improve the modelled spectra to better describe the observations.

\subsection{Stratified Abundance Model} \label{stratified}

\begin{figure*}
    \includegraphics[width=\textwidth]{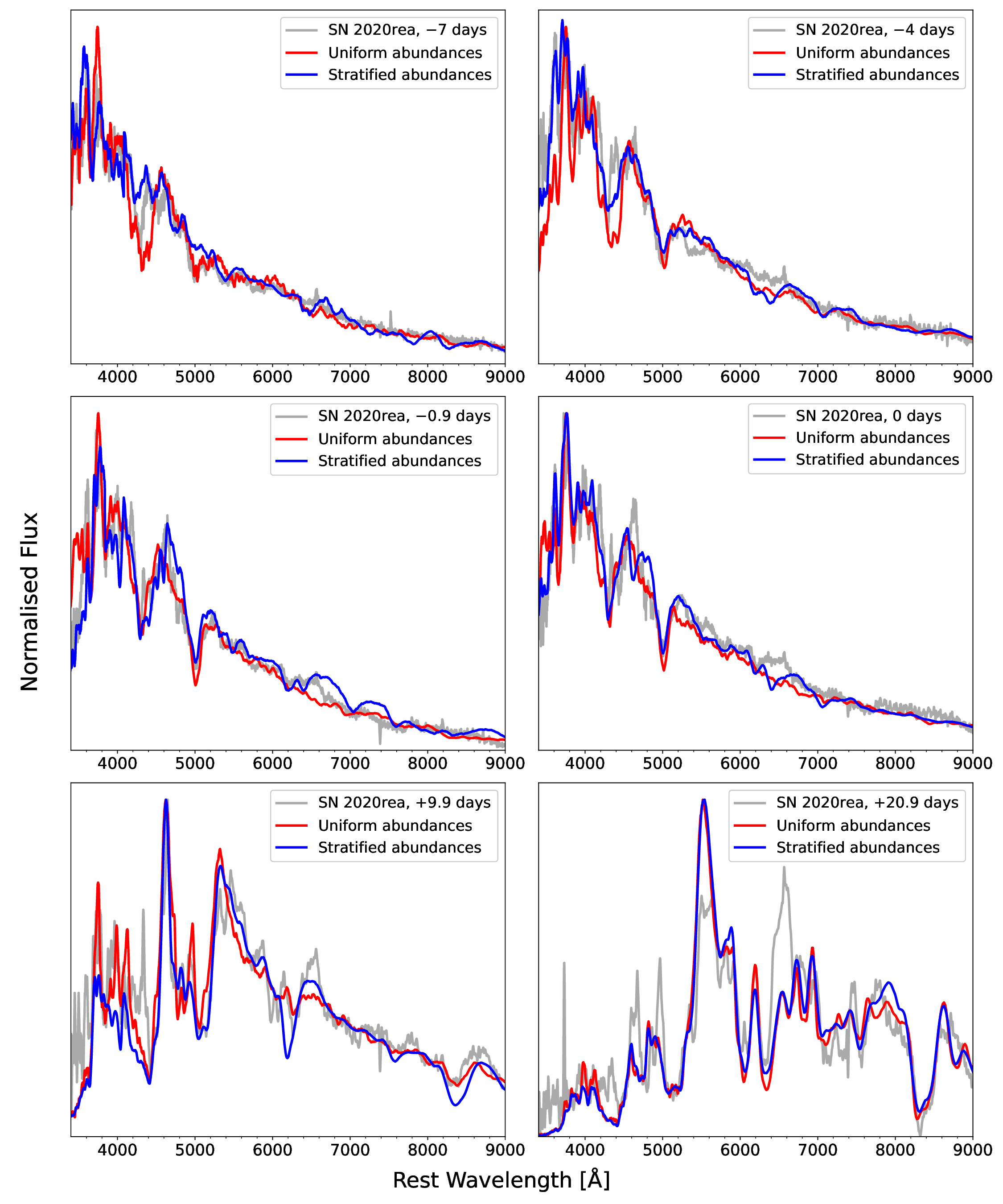}
    \caption{The observed spectra of SN~2020rea obtained between $-$7 and +20.9 days (w.r.t. $g_\text{max}$) compared to the best-fitting \textsc{tardis} models assuming stratified and uniform abundance profile.}
    \label{strat_spectra}
\end{figure*}

As the SN evolves with time, the ejecta expands, and the photosphere recedes into deeper layers, meaning that the earlier epochs' spectra probe the outer regions. In contrast, the later epochs probe the inner regions of the ejecta. This allows us to construct a velocity-dependent abundance profile. Thus, the parameters taken from the best fit uniform abundance model (section~\ref{unifrom}) for each epoch were condensed to build the stratified model \citep{Stehle_Mazzali_Benetti_Hillebrandt_2005, barna_abundance_2017}.

The \textsc{tardis} models for the six epochs using a stratified abundance profile are shown in Figure~\ref{strat_spectra}. The mass fractions are varied in a velocity grid between 5800--6900 km s$^{-1}$ with a step size of $\sim$ 200 km s$^{-1}$. For a comparison, the uniform abundance models are also shown.

The velocity-dependent, stratified abundance structure of the \textsc{tardis} model has been illustrated in Figure~\ref{mass_fracs}. A very low amount of Fe is used for the synthetic spectra at pre-maximum and maximum epochs, which is then increased for post-maximum epochs. Thus, the higher inner velocities have negligible traces of Fe, which then increases when the SN enters the Fe-dominated phase at later epochs. The mass fraction of $^{56}$Ni has been changed over the velocity range based on \cite{barna_type_2018} template. Oxygen has been used as a `filler element', as varying this leads to no significant difference in the synthetic spectra. The presence of Mg in higher velocity outer layers is possibly because a higher amount of Mg exists in the outer regions of the progenitor star. Mass fraction of the IMEs such as Ca, C, Si and S are nearly constant. This implies that the abundance of these elements does not change much as the SN evolves.

\begin{figure*}
    \begin{subfigure}
        \centering
        \includegraphics[width=\columnwidth]{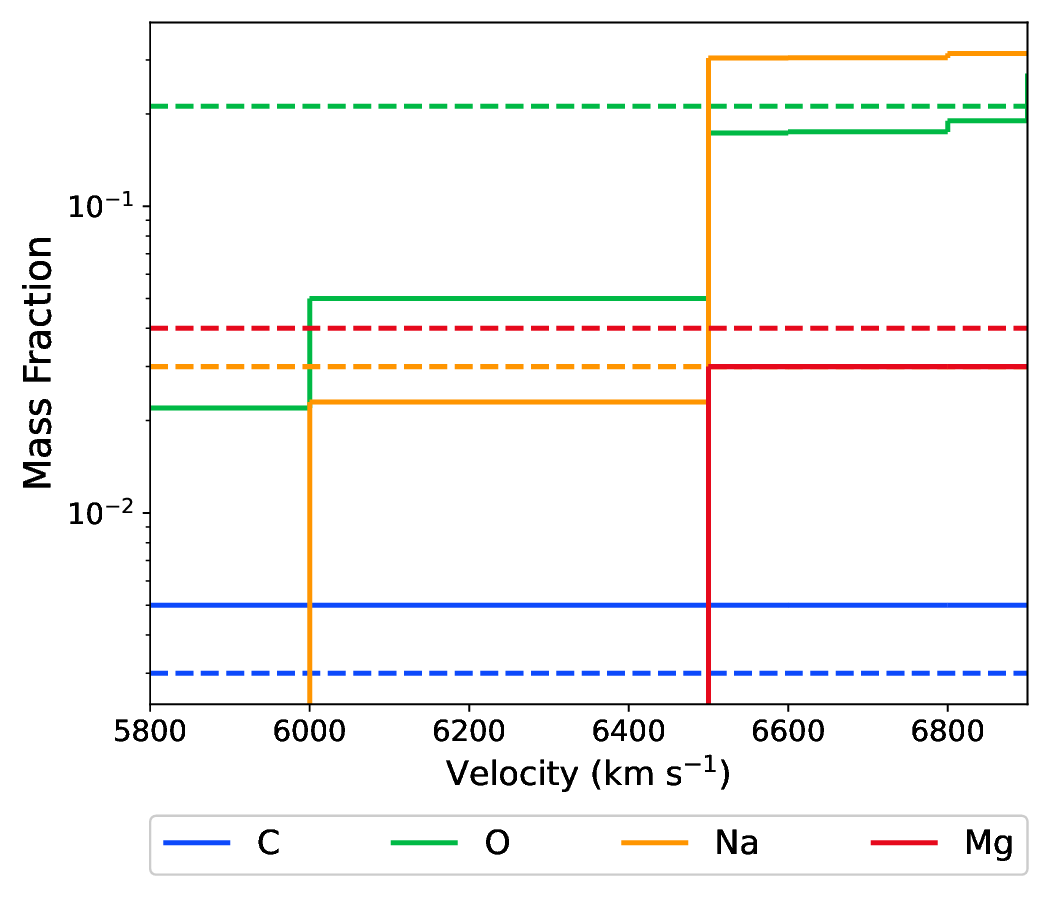}    
    \end{subfigure}
    \begin{subfigure}
        \centering
        \includegraphics[width=\columnwidth]{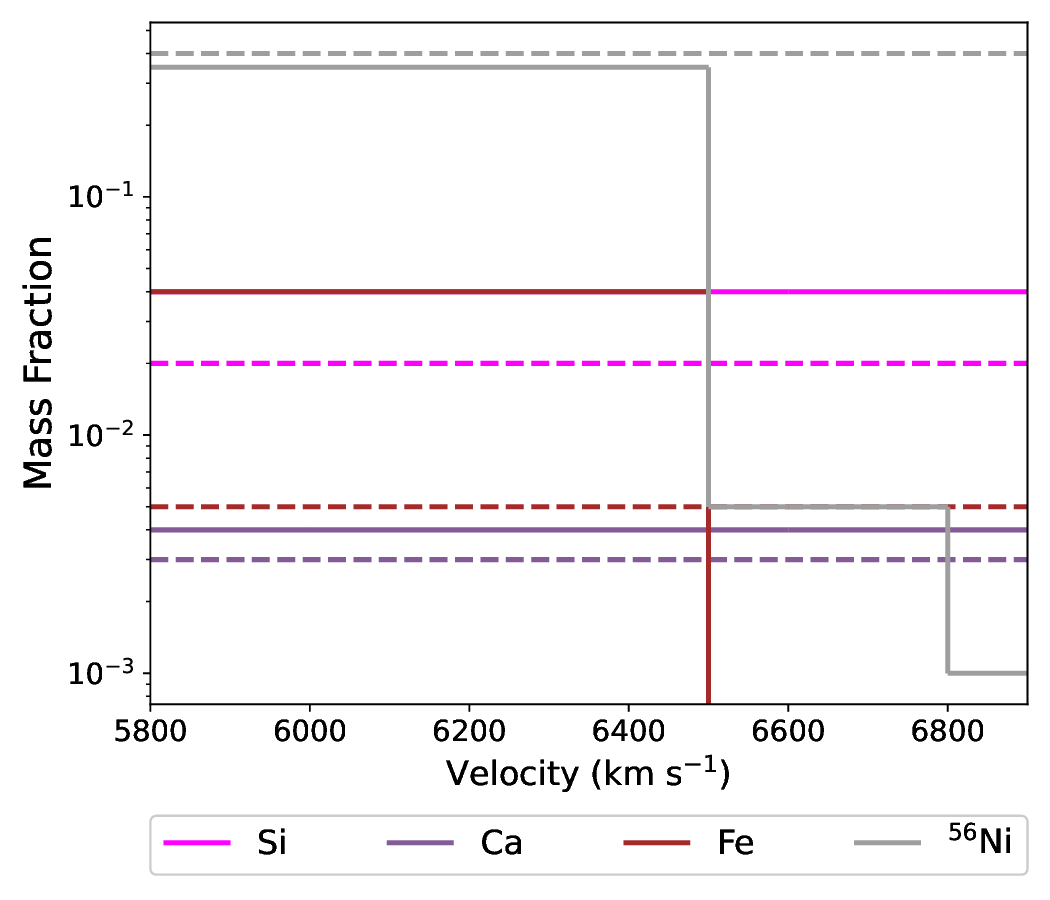}    
    \end{subfigure}
    \caption{The best-fit chemical abundance structure for the synthetic spectrum. The dotted lines show the best-fit abundance values used in the uniform abundance model.}
    \label{mass_fracs}
\end{figure*}

\begin{figure*}
    \includegraphics[width=\textwidth]{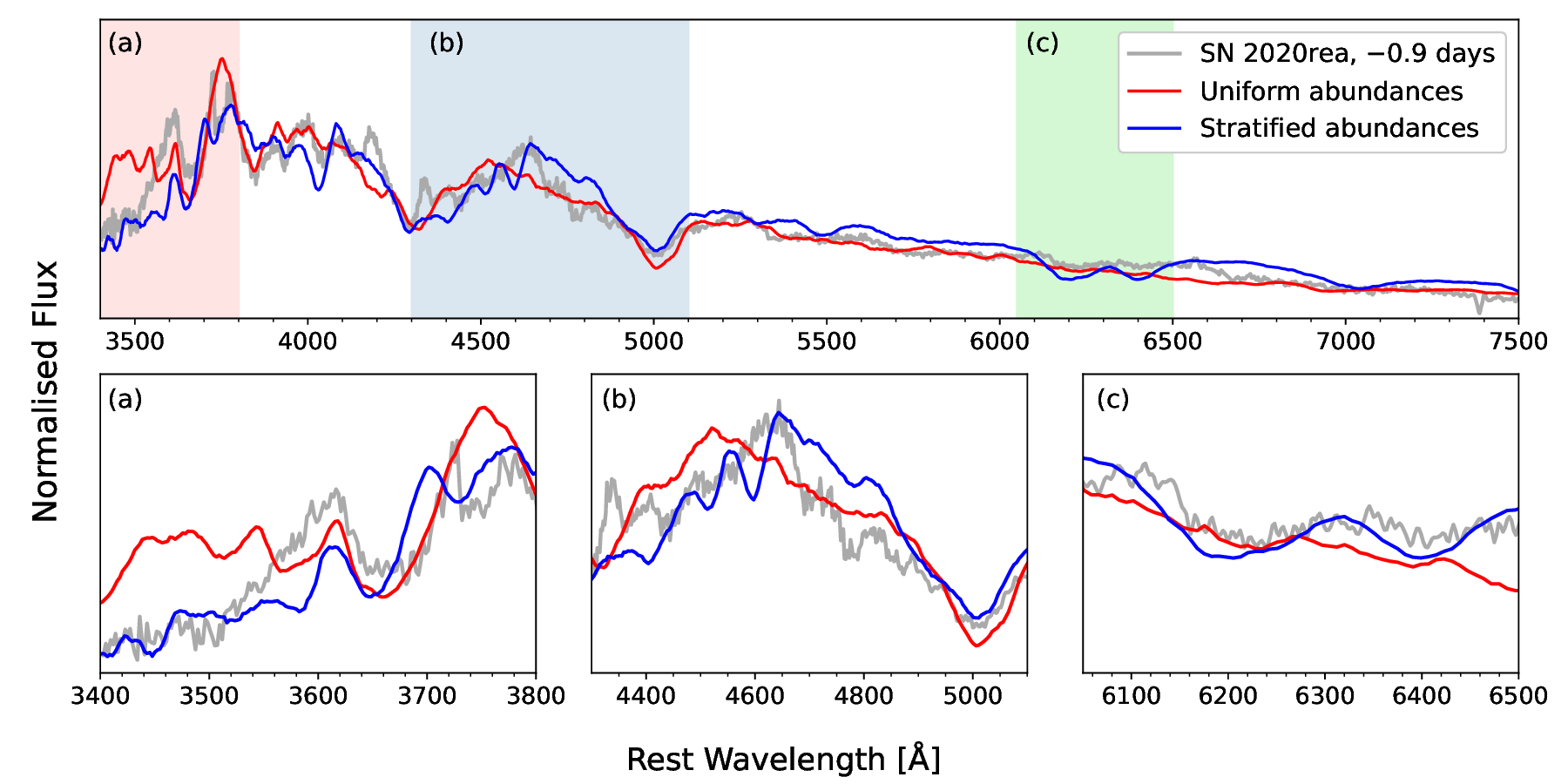}
    \caption{Best-fit modelled spectra with uniform and stratified abundances compared with the observed spectrum of SN~2020rea (at $-$0.9 days w.r.t $g_\text{max}$). The shaded bars highlight the (a) 3400--4200 \angstrom\,region where the stratified abundance model provides a better fit, (b) the peaks at $\sim$ 4700 \angstrom\,and the (c) ``W'' feature  $\sim$ 6000 \angstrom\,being reproduced, which were missing in the uniform abundance model.}
    \label{zoom}
\end{figure*}

$^{56}$Ni plays an important role in the innermost region, causing around half of the line emissions in the synthetic spectra \citep{barna_type_2018}. This value, however, decreases for the outer regions of the ejecta. For the pre-maximum epochs, the stratified abundance model better fits 3400--4200 \angstrom\,region (Figure~\ref{zoom}a). The peaks at around 4700 \angstrom\,for near-maximum, which were earlier missing in the pure-deflagration model, are also present in the stratified model (Figure~\ref{zoom}b), representing the presence of Ti II in near-maximum spectra. The ``W'' feature observed at around 6000 \angstrom\ in the spectra originates from S during the initial evolutionary phases. As the SN enters its Fe-dominated phase, this S is converted into Fe \citep{singh_optical_2022}. This model has reproduced it (Figure~\ref{zoom}c). However, the stratified abundance model could not accurately reproduce the 6000--7000 \angstrom\,region in the +20.9 day spectrum. This could be because the inner regions have a more uniform abundance profile, as explained in a deflagration scenario. At the same time, the outer layers are more stratified, hence the accuracy of the stratified model in the pre-maximum and near-maximum phases. 

In the pre-maximum spectra, Fe III features near the 4000--5000 \angstrom\,band have been noted clearly. The C II feature near 6580 \angstrom, prominent in fainter SNe, is very weak here, as SN~2020rea is a bright SN \citep{barna_sn_2020}. Thus, overall, the pre-maximum and maximum spectroscopic features obtained in the synthetic spectra are typical of a bright SN Iax. The velocity profile also matches that of bright SNe Iax. 

As the ejecta expands, the velocity of the layers increases, thus resulting in lower inner velocity described by later epochs, as the inner velocity goes inwards. The outer velocity of the model ejecta remains constant, thus increasing the size of the line formation region.
Hence, abundances in inner layers are derived from models of later-phase spectra. The outer regions of the ejecta seem to be stratified rather than mixed, based on the better fit of this model than the uniform model, especially for the spectra obtained at the early epochs. Therefore, this difference in the abundance distribution could indicate that the propagation of the outer layers of the ejecta is different from a pure deflagration scenario.

\section{Conclusion} \label{conclusion}

SN~2020rea is categorised as one of the most luminous members of the Type Iax SNe family with a peak absolute magnitude of $M_\text{V, peak} \sim {-18.30}\pm{0.12}$ mag. We have presented a comprehensive study of the SN~2020rea spectra at six epochs from $-$7 days to +21 days since peak brightness. The primary goal of this study is to probe the nature of the SN ejecta by mapping the distribution of various elements in it using the abundance tomography method \citep{Stehle_Mazzali_Benetti_Hillebrandt_2005, barna_abundance_2017}.

\cite{singh_optical_2022} performs a thorough photometric and spectroscopic study of SN~2020rea, and compares it with the other significant SNe in the sub-class. They find the post-peak decline of the pseudo-bolometric light curve to be slower than typical light curves predicted by deflagration models. Spectroscopic analysis performed in the paper shows a higher Fe line velocity than Si line around maximum light, which indicates mixing of fully burned material \citep{phillips_peculiar_2007}. 
They conclude that the pure deflagration model of a WD looks to be the most promising explosion scenario for SN~2020rea as it predicts a well-mixed ejecta.

In this study, we have used the 1D Monte Carlo radiative transfer code \textsc{tardis} to perform spectral modelling. Assuming uniform elemental abundances throughout the ejecta, spectral modelling was performed for all six epochs. However, in the pre-maximum and near-maximum spectra, certain spectral features could not be reproduced, which led us to consider a scenario where the ejecta is not fully mixed, especially for earlier epochs. After analysing and comparing the synthetic spectra obtained using the uniform abundance model, the parameters for the best fit for each epoch were used to build a stratified abundance model. This velocity-dependent, stratified abundance structure was adopted from \cite{barna_abundance_2017}.

A comparative analysis of both the models and the observed spectra was performed, from which we concluded that the stratified abundance model was overall a better fit, especially for the pre-maximum epochs. Thus, the ejecta's inner layers are found to be more homogenous, while the outer layers are found to be stratified in structure. These findings suggest that, with time, the SN ejecta transition from being stratified to becoming uniformly mixed, challenging the pure deflagration interpretation. The explained ejecta structure is consistent with expectations from WD explosion physics and well-established explosion models \citep{khokhlov-1991, roepke_three-dimensional_2007, seitenzahl-2012, fink_three-dimensional_2014}. The weak deflagration models generally produce a well-mixed inner ejecta with only partial stratification in the outer layers, whereas the delayed detonation model produces a highly stratified ejecta composition by suppressing large-scale mixing \citep{sim-2013, fink_three-dimensional_2014}. Detailed modelling covering the full spectral time series of multiple SNe Iax is required to constrain their possible explosion scenario and progenitor system.

\section*{Acknowledgements}
We sincerely thank the anonymous referees for their insightful and constructive comments that have significantly enhanced the manuscript.
We acknowledge the support of the NIUS programme of HBCSE-TIFR funded by the Department of Atomic Energy, Govt. of India (Project No. RTI4001). MS acknowledges the financial support provided under the National Post Doctoral Fellowship (N-PDF; File Number: PDF/2023/002244) by the Science \& Engineering Research Board (SERB), Anusandhan National Research Foundation (ANRF), Govt. of India. KM acknowledges the support from the BRICS grant DST/ICD/BRICS/Call 5/CoNMuTraMO/2023 (G) funded by the Department of Science and Technology (DST), India. This research made use of \textsc{tardis}, a community-developed software package for spectral synthesis in SNe \citep{kerzendorf_spectral_2014, kerzendorf_tardis-sntardis_2023}. The development of \textsc{tardis} received support from GitHub, the Google Summer of Code initiative, and from ESA's Summer of Code in Space program. \textsc{tardis} is a fiscally sponsored project of NumFOCUS. \textsc{tardis} makes extensive use of \textsc{Astropy} and \textsc{Pyne}.

\bibliography{references}
\balance

\end{document}